\documentclass{elsart}
\usepackage{epsfig}

\usepackage{amssymb}

\newcommand{\refe}{}

\begin{document}

\begin{frontmatter}

\title{Flux Modulation from Non-Axisymmetric Stuctures in Accretion Discs}

\author[label1]{Peggy Varni\`ere \& Eric G. Blackman}
\address[label1]{Department of Physics and Astronomy and Laboratory for Laser Energetics,  University of Rochester, Rochester NY, 14627}

\begin{abstract}
Non-axisymmetric accretion discs can lead to flux variability.
Here we provide an analytic framework for
modeling non-axisymmetric structures like hotspots
and spiral waves and their influence on  observed timing measurements.
The presently unexplained Low-frequency Quasi-Periodic Oscillations (LFQPO),  
observed in X-ray binaries and cataclysmic variables, 
could be the result of such discs. 
Our framework serves as a guide to quantify 
the properties that non-axisymmetric structures produced by 
nonlinear accretion disc models must have in order   
to explain observed features such as  LFQPOs. The results
from our microquasar appliations
also provide analogous predictions for X-ray modulation 
in active galactic nuclei. The formalism and physical interpration 
is of practical use for generic non-axisymmetric accretion disc systems.
\end{abstract}

\end{frontmatter}

\section{Introduction}
\label{sec:Intro}
X-ray emission from accreting black holes in binary stellar systems
varies on time scales ranging from milliseconds to years. Variability
on time scales longer than days appears to be driven by changes in the 
accretion rate onto the black hole, and is often manifested as transient 
outbursts in which the luminosity of a source changes by a million-fold 
Lasota (2001).
At the shortest time scales, quasi-periodic oscillations (QPOs) are observed 
in the X-ray emission. The highest frequency QPOs ($> 100$Hz)
are consistent with those expected 
from general relativistic orbits near the innermost stable orbit around the 
black hole, and they are likely caused by inhomogeneities in the inner 
accretion flow (Remillard et al., 2002, Stella \& Vietri, 1998). 
However, the cause of low frequency  QPO (LFQPO)s {0.1--20~Hz} is still a 
mystery.  This motivates the work herein.
{\refe For neutron star X-ray binaries, low frequency QPOs
can be sub-divided into more precise categories (e.g.  Psaltis et al.,1999).  
Here we focus on a basic paradigm for the modulation
and leave explanations of possible harmonic relations
and phase lag behavior Varni\`ere (2005) for further work.
Black hole microquasars avoid the role of any solid stellar surface, 
and offer a purer probe of the accretion  process
without interaction with a stellar surface.  The 
 basic properties of LFQPOs that any model needs to explain are these:}

{(1)} {LFQPOs 
often appear to be present in widely spaced observations, 
over a period of 
weeks,  with fractionally narrow frequency widths
($\Delta \nu / \nu \approx 1/30$)} (Morgan et al., 1997, 
Remillard et al., 1999). 
{\refe Although the observations
are not continuous, the fact that the LFQPO are
 observed to be the same (within a
very small variation) over several observations within the same
week/month, suggest that the QPO are likely always present 
during that time, at least when averaged over such long time scales within
the hard state;  there are periods where the LFQPO appear and 
disappear on timescale of a few seconds.}
If LFQPOs result from inhomogeneities orbiting {\refe at a Keplerian
speed} 
in the accretion disc they are constrained to a narrow range of radii. 

{(2)} 
LFQPO  amplitudes are  typically 5-10\% RMS (root mean squared), 
but can reach 20\% RMS. 
Given that the X-ray luminosities of  black hole binaries approach 
$10^{39}$ erg s$^{-1}$, LFQPOs involve an enormously energetic fraction of 
the accretion flow. 

{(3)} On the other hand, LFQPOs 
are transient features, so the 
spectral properties of the X-ray emission when LFQPOs are present and absent 
can be used to constrain the LFPQO origin (Muno et al., 1999, 
Sobczak et al., 2000). 

{(4)} { The 0.1--20~Hz LFQPOs only appear 
when the non-thermal component of the X-ray spectrum is strong, and their
fractional amplitudes increase with energy between 2--20~keV}.
However, a
{ thermal component} that contributes $\sim 10\%$ of the X-ray emission at 
lower energies must also be present, and {the frequencies of the LFQPO appear to be correlated best with this}.  
The energy output of accreting 
black holes can generally be decomposed into a $\sim 1$~keV thermal 
component (thought to originate from the optically thick accretion disc) 
and a non-thermal component with a power-law spectrum extending beyond 
100~keV (thought to result from  inverse-Compton scattering of 
cool photons by a corona of hot electrons.) 
Thus, it appears that the 
0.1--20~Hz QPOs either originate in the boundary between the disc and 
the corona, or are part of the mechanism which accelerates the hot 
Comptonizing electrons. 

{\refe As mentioned,  LFQPOs can be further subdivided
(Psaltis et al., 1999), but 
the above represents a basic set of characteristics
that a zeroth order LFQPO flux modulation model should explain.}

Two different conceptual paradigms for LFQPOs 
have been proposed in this regard:
(1) In the centrifugal pressure supported boundary layer model (CENBOL,
Chakrabarti \& Manickam, 2000), a QPO is produced by a 
shock in the accretion flow where it makes a transition from a 
Keplerian disc to a hot Comptonizing region. If the cooling time of 
the post-shock region is resonant with the free-fall time at the shock, 
the shock can oscillate radially with a frequency on order a Hz. The 
QPO is, in principle, produced because the shock modulates the flux of seed photons 
that reach the Comptonizing post-shock region.
In this model, the LFQPO would 
represent a global radial oscillation that modifies
the emitted flux. 
(2) {\refe A second paradigm for
LFQPOs is non-axisymmetric structures in the disk.
Their motion around the black hole would lead to flux modualtion.
These structures could move at the Keplerian speed,
a precession speed (such as Lense-Thirring) or a phase
velocity associated with a structural instability.
One example occurs in }
the accretion-ejection instability model (AEI, Tagger \& Pellat, 1999,
Varni\`ere \& Tagger, 2002). Here a spiral shock forms in 
an accretion disc threaded by a vertical magnetic field.
The dispersion relation is similar to that of a galactic disc but with the 
gravitational potential replaced by a magnetic one. 
At the co-rotation radius, at which the Keplerian velocity matches 
that of the pattern, a Rossby vortex 
forms and emits vertical Alfv\'en waves into  a corona. 
The LFQPO would result from the orbit of this non-axisymmetric
structure in the disc. 2D numerical simulations of this instability, for example,  demonstrated (Varni\`ere {\em et al.}, 2003) that the X-ray flux was
modulated. 
However, in 2D,  the constraints on the cause 
of the modulation and how it can be stronger and weaker were limited.

{\refe In this paper, 
we focus on the paradigm exemplified by the latter
type of model, namely the production of LFQPOs by the orbit of
a non-axisymmetric structure. Here we 
do not focus on  the specific origin of the structure but rather investigate
the consequences of the presence of such a structure.}
We investigate the specific question
of whether a stable pattern in an accretion disc can 
reproduce the observed characteristics of LFQPO simply 
by its motion around the central engine. 
In Sec. 2 we give analytical formulae 
that can be used to model arbitrarily shaped
blobs and spirals as non-axisymmetric disc features.
In Sec. 3 we compute the emission from
an accretion disc with such  non-axisymmetric
structures and discuss
 the influence of the shape of the particular non-axisymmetric  
structures on the RMS LFQPO amplitude.
We also  discuss the parameter choices which provide
the most observationally consistent non-axisymmetric structures for the microquasar example discussed above. We also note that  
the formalism is generically applicable to other 
non-axisymmetric accretion disc systems.
We conclude in Sec. 5.

\section{Modeling non-axisymmetric discs}

Blobs and spirals are two useful categories of non-axisymmetric structures.
We use blob to indicate a generic   localized
feature and spirals to indicate the more specifically identifiable
global structure produced by gravitational or MHD instabilities--the latter
via the accretion-ejection instability (AEI) 
i.e. Tagger \& Pellat 1999.
Here we do not detail the formation of these non-axisymmetric
structures but  focus on providing an analytic framework
that characterizes their shape for practical use
and allows observational implications for flux modulation to be quantified.

Because discs of observed systems such as microquasars and
active galactic nuclei (AGN) are not resolved, the  
disc structure cannot be directly imaged. But from studying the timing evolution
of the flux, especially the presence of 
the LFQPO with its frequency and RMS amplitude, we can 
infer what structures might be present.
We concentrate on the RMS amplitude of the modulation, 
assuming that the frequency is already matched by the presumed location 
of the given structure.

Taking into account the disc thickness is a particularly important 
aspect of our endeavor.  
For discs viewed at highly inclined angles,  the shadowing 
      from a local thickening of the disc can be important.
The height and temperature profiles 
are coupled in the hydrostatic equilibrium approximation. 
Although we make this approximation here,
detailed $3$D simulations will ultimately be needed to
get an exact profile of $h$ and $T$ for a more realistic disc.

\subsection{Analytic expression for the disc thickness}

For the disc thickness, we write 
\begin{eqnarray}
h(r,\phi) = h_o(r) 
+ h_1(r(\phi),\phi))
=h_o(r) + s(r-r_s)d(r)
\label{1}
\end{eqnarray}
where $h_1$ is a perturbation in thickness around the unperturbed
thickness $h_o$, $r$ is the radial location, and $\phi$ is the azimuthal
angle.
We have used $h_1=s(r-r_s)d(r)$, 
where $s$ is  a ``shape'' function.
The latter is 
finite only near the disc structure causing the non-axisymmetry.
The blob or spiral wave feature is  localized 
by $r_s \equiv r_c e^{\alpha(r) \phi}$, where $r_c$ is the
point where the structure  begins and $\alpha$ is the opening angle 
of the structure. The quantity  $d(r)$ is a thickness function which is 
defined as the height of the disc at each point.

We now      take $s(r-r_s)$ to be a Gaussian and $d$ to be 
a power-law in $r$. This provides a simple 
but useful framework to model  non-axisymmetric structures.
From (\ref{1}) we then have 
\begin{eqnarray}
h(r,\phi) = h_o(r) + {\tilde \gamma} \left(\frac{r_c}{r}\right)^\beta
                 e^{-0.5 \left(\frac{r-r_s}{\delta}\right)^2}, 
\label{2}
\end{eqnarray}
where the constant $\beta$ measures how fast the 
    thickness decreases from the maximum,
$\delta$ parameterizes  the radial extent of the structure,
and $\tilde \gamma$ 
defines the disc thickness at $r_c$, 
We allow the maximum height $\tilde \gamma$ to be a function of the 
unperturbed 
thickness at that point, that is ${\tilde \gamma}=\gamma h_o(r_c)$. 
This allows consideration of cases with similar $h_0/r$ but different $r_c$.
We also consider the number of times the
non-axisymmetric structure winds around the disc.

The influence of the parameters in (\ref{2}) is illustrated in Fig. 1.
The role of $\alpha$ is seen in Fig. 1a which shows 
a  top view of  two spirals'  ``spines''  
(i.e. the line defined by $r_s(\alpha,\phi,r_c)$ tracing  
 the maximal height above the unperturbed disc  at each $\phi$.). 
The larger the  $\alpha$ the more open the spiral.
The role of parameters ${\tilde \gamma}$, $\delta$, and $\beta$
are seen in Fig. 1b which 
shows cross sections of the disc height profile for different $\phi$
projected into the 
$r,z$ plane for $r_c=1.5$.
Increasing $\delta$ would increase the 
full width-half-maximum of the peaks, which
in practice has a larger effect on the rise 
to the peak than the fall because the Gaussian
perturbation is superimposed upon a positively sloped disk.
Increasing $\tilde \gamma$ would increase the maximum height above
the $h_o$, and the downward slope of the line that connects the peaks 
at different $\phi$ values would be steeper for larger $\beta$.
Fig. 1c
 shows a 3-D close-up of the inner
region of the disc with a one-armed spiral characterized by $h_o/r = 0.01$, 
$r_c=3$, $\alpha = 0.07$, ${\tilde \gamma} = 0.3$, $\beta = 2$, and 
$\delta = 0.2$ at an viewing angle of $70^{\circ}$ from the normal.

\subsection{Emission}

Having obtained an expression for the disc height,
we compute the temperature using the approximation that $c_s = h\Omega$,
    which gives $T=\mu/R_{gas} c_s^2$. This means, for example, 
that a change in temperature by $20\%$
    is related to a change in thickness of about $45\%$.

In order to determine the observational effect of a non-axisymmetric disc
thickness, we compute the flux as a function of azimuth 
that an observer would received as time-dependent 
flux modulation when the disc rotates.
Assuming that the spectrum from each point is a blackbody,
we use the height-temperature relation above to  compute the photon flux
\begin{equation}
f(E) = {{2} \over {c^2 h^3}} {{E^2} \over {\exp[E/kT] - 1}}.
\end{equation}
We then sum the flux from each cell, multiply by 
$cos([\pi/2 - \theta] - \zeta)$, where $\theta$
is the disc inclination angle from the normal
and \begin{equation}
\zeta = atan\left({{dh} \over {dr}} \cos \phi - {{dh} \over {r d\phi}} \sin{\phi}\right).
\end{equation}
We trace rays from the individual cells to the observer in order to 
determine whether the flux is intercepted by the outer portions of the disc.
We compute the observed spectrum as the structure 
rotates by viewing a single
snapshot at all angles $\phi$. 
The spectrum is taken as multi-temperature blackbody, 
and we sum the number of photons received. 
We compute the Fourier transform of 
the profile in order to estimate  the amplitudes of any modulation at 
multiples of the pattern frequency.

{\refe Our simple spectral modeling does not include
scattering in the disc atmosphere, the presence of an electron
corona, or special and general relativistic (GR) 
effects in part because the LFQPOs observed in microquasars
could be coming from shadowing by structures at $> 100$ gravitational radii.
Also, the non-relativistic formalism applies at all radii for
generic disks systems around stars.
We note however, that non-Keplerian origins of the non-axisymmetries 
leading  to LFQPOs  are also possible, such as  
binary induced precession, spiral wave phase velocities that differ from 
Kelperian, or Lense-Thirring precession, 
The latter is  an intrinsically general relativistic
phenomenon. While relativistic effects should be considered
in future more detailed applications to compact systems, 
here we focus on qualifying the effect of the key 
parameters in (\ref{2}) on the modulation, not on the particular
source of the the non-axisymmetry.}

\section{Simulated flux modulation}

      Table \ref{tab:simul_modulation} shows 
      a subset of our numerically solved cases for the blob and spiral.
Cases 1-15 can be considered spirals, whilst cases 16-19 can be
considered blobs because their structures extend less than $2\pi$ radians.
For all cases we have taken
an inclination angle of $70^{\circ}$, motivated by
the inferred observation angle of GRS 1915+105 (Mirabel \& Rodríguez, 1994).

There is degeneracy in the relative influences
of the various parameters of (\ref{2}) on 
the RMS amplitude of modulation of Table 1, but it  
is instructive to discuss the influence of each of the  
parameters separately.

\subsection{Spirals and physical interpretation of the geometric parameters}

      First notice the lack of influence of the range of 
      $\phi >2\pi$ in Table 1: For each $\phi$, most of the modulation
      comes from  the most inward ``bump'' in $h(r)$.
      Allowing for more than 1 bump
      does not modify the RMS modulation significantly,  
      as seen in cases  $\# 1, \#2, \#3$. 
      For the majority of the other cases 
      we will therefore consider only the range of $\phi \le 2\pi$.

      The parameter  $\tilde \gamma$, measures  the maximum thickness of the perturbation. 
      Its effect  is  evident in comparing cases $\# 3,\# 6,\# 7$. 
      The greater $\tilde\gamma$, all else being equal, 
      the stronger the modulation.  This is because a thicker 
      structure  more strongly shadows the inner disc.  A change in thickness of $70\%$ means
      a temperature change of  $\sim 50\%$. This somewhat extreme case is taken
      to illustrate the role of $\tilde \gamma$ on the RMS amplitude. 
      Constraining the amplitude modulation from first principles 
      requires a $3$D MHD disc simulation that exhibits a 
      spiral wave instability. Here the non-axisymmetry would emerge rather 
than be  imposed. Previous $2.5$-D (i.e. no dynamical vertical structure) 
disc simulations unstable to the AEI exhibit spiral waves, 
      and lead to a  temperature variation typically of order of $20-30\%$ 
      in the hydrostatic equilibrium approximation (Varni\`ere et al,  2003).  
      But in 3-D, hydrostatic
      equilibrium would underestimate the perturbation thickness
of an AEI generated spiral
      because of additional heating from spiral shock dissipation.
Thus  a  higher $\tilde \gamma$ than that
      obtained from the $2.5$D simulation can be expected, 
provided the AEI survives in 3-D. 
      
      Because ${\tilde \gamma}\equiv \gamma h_o(r_c)$ is 
the maximum height 
      of the spiral arm, a
      change in $h_o$ (the zeroth order disc thickness)
      also increases the modulation.
      By comparing cases $\# 14$ and $\# 15$ we see that changing $h_o$ by a factor 
      of two is not exactly  the same as changing $\tilde \gamma$ by a factor of two
because increasing $\tilde \gamma$ also
      increases the difference between the maximum height of the perturbation
and $h_o$.

      As discussed at the end of Sec. 2, the parameter $\beta$ measures how fast the maximum height 
      along the spiral decreases with radius.
Its effect is revealed in cases
      $\# 3, \#4, \#5$.  The RMS amplitude of the modulation
      increases with $\beta$ for an $\alpha > 0$ spiral. 
A rapidly decreasing height perturbation with radius 
means that the height also rapidly decreases along the spiral. This 
causes a stronger modulation by producing a stronger 
azimuthal variation.
This is particularly important for tight spirals
(small $\alpha \sim 0.05$): were it not for a large $\beta$,
little non-axisymmetry would otherwise arise.

The simulations relevant for studying  
      the influence of 
      $r_c$,  the initial radius where the perturbation begins $(r_c)$, 
are $\#6$,
      $\#9$, $\#10$ and $\#11$. We see a maximum in the RMS amplitude for $r_c$
      around $2\ r_{in}$, where $r_{in}$ is the inner disc radius: When the spiral is too near $r_{in},$ it obscures
      less of the inner disc which is the most luminous part. 
On the other hand, if the spiral is too 
      far out, it can only obscure a less luminous outer region 
giving a smaller modulation. The combination of these effects
is an intermediate $r_c$ for maximal modulation.
Since the position of the spiral 
is chosen based on the LFQPO frequency, 
a correlation between the RMS amplitude 
and the frequency of the QPO is expected.
In practice this may be  hard to disentangle from
 variations in the other parameters.

The influence of the parameter $\alpha$, which determines the 
opening of the spiral wave can be seen from simulations  
      $\#6$, $\#12$ and $\#13$. There we  Ssee that the more open the spiral,  
      the higher the RMS amplitude.  This is 
because a more  open spiral means a more   
     non-axisymmetric thickness profile, leading to a 
higher RMS amplitude of modulation.

Finally, consider the parameter $\delta$, which measures
the width of the spiral or blob at its base.
From cases $\#6$ and $\#8$ we see that
a larger $\delta$, 
implies a larger RMS amplitude.  
For our choice of a highly inclined system, this trend 
results because a larger $\delta$ makes more of the region inner to the peak
of the perturbation more perpendicular to the line of sight. 
This produces   
a larger observed flux when the observer is looking at the disk
from an azimuth for which the line of sight intersects  
the inner part of the perturbation.
But as the disk rotates, the outer edge of the perturbation comes
into view, and the shadowing of the inner region occurs similarly
for large or small $\delta$.  The contrast in flux 
(and thus the RMS amplitude of the modulation) is thus
larger for larger $\delta$,  explaining the trend.

\subsection{Blobs or hot spots}

Because the difference between spirals and blobs in our model is 
just whether the structure extends for a range of $\phi \ge 2\pi$ (spiral)
or $\phi < 2\pi$ (blob), 
the parameter influences are similar for the two cases.
The relevant blob simulations are 
$\#16$, $\#17$, $\#18$, and $\#19$. 
There we see that a blob can create a non-negligible 
modulation of the X-ray flux but to reach a large RMS amplitude, 
the blob needs a large azimuthal extent--making
it more banana or spiral shaped. The reason is that 
a localized structure does not provide much shadowing
over a disc rotation period and thus offers only a weak RMS modulation. 
In contrast to spirals, such blobs probably cannot account for observed
LFQPOs in microquasars.

\section{Expected dependence of RMS amplitude on inclination and energy}

Because the modulation comes from shadowing, 
the disc inclination angle $\theta$ 
is important in determining the maximum shadowing
and thus the maximum 
RMS amplitude that a given choice of parameters can produce.
Motivated by a comparison with the LFQPOs of  GRS1915+105  
(see {\em e.g.} the review Mclintock \& Remillard 2004), 
we again focus on the parameter choices of simulation $\# 8$ and vary the
inclination angle to obtain different values of
the RMS amplitude. 
Fig. \ref{fig:incli} shows 
the result.
As expected, the more edge-on the view, the higher the RMS amplitude.
       Present observational  data  are insufficient
       to definitively confirm or contradict the predicted behavior.
       Such a trend  could explain the {\bf  extremely weak} 
LFQPO in Cyg X-1, as the RMS amplitude
       expected from the inferred inclination angle 
is very small\footnote{Cyg X-1 seems to show a weaker and
       broader structure than the ``usual'' LFQPO that could be a weak
        LFQPO in aggrement with this simple model.}. 
	More objects with a wider range of inclination angles are needed.
	Additional techniques of
	determining the disc inclination angle in microquasars
	besides  using the jet propagation direction 
	would also be desirable. 
{\refe In the case of neutron stars, data from 
eclipsing binaries and dippers provides good information
on the binary and thus disk inclinations (e.g Frank et al., 1987).}

{\refe It is important to note that the dependence of the rms amplitude on the
inclination angle shown in Fig. \ref{fig:incli} is displayed only for the case 
in which all  other parameters are identical. 
Therefore the trend shown should be seen inside the same
``family'' of objects--those having similar disk properties. 
It was shown in table \ref{tab:simul_modulation}
that some parameters can have a greater 
 influence on the rms amplitude than the inclination. These
paramaters could give rise to low inclination source with a strong spiral 
having a higher rms amplitude than a high inclination source with 
a very small spiral}

       Using the relation between $T$ and $h$, we can also
       study how the RMS amplitude behaves as function
       of energy. Several studies show that LFQPO amplitudes rise 
with energy up to 15-20keV, although the behaviour at higher energies
 seems to depend on the source, for example in GRS $1915$+$105$
       (Tomsick \& Kaaret, 2001). 
       We have not included a corona (high energy 
       component) in our simple spectral modeling herein,  
but we can make a qualitative prediction of the LFQPO amplitude vs. energy
trend by studying the RMS amplitude as function of our predicted 
       disc flux, supplemented by knowledge of the 
disc/corona flux ratio taken from observation. 
We find that the RMS amplitude of the modulation gets bigger at higher energy.
his is simply because a spiral with $r_c/r_{in}\sim 2$
is hotter than the surrounding disc 
and therefore its blackbody contribution to the spectrum peaks at a higher 
temperature. When a corona is present, 
this trend would apply at low
       energies, but we expect a critical energy
above which the corona strongly dominates and the trend reverses.

{\refe We emphasize that in 
order to predict the exact position of the modification in
the trend at high energies, 
we need to take into account not only the disk but
also the corona self-consistently. This
is beyond the scope of the present paper which is intended to
show the basics of how flux can be modulated by non-axisymmetric structures.}

\section{Summary and Discussion}

We have presented a simple  analytical framework
for modeling and interpreting how  non-axisymmetric disk features such
as spirals or blobs  can create modulation of the observed flux from an accretion disk.
We connect the  disk geometry to the flux 
via the hydrostatic equilibrium assumption.
Our focus on spirals is motivated by their production 
in disc simulations of the AEI and the efficacy
with which such spirals seem to account for the observed
LFQPO properties of microquasars, see {\em e.g.} the review 
McClintock \& Remillard, 2004. 
Our formalism does not depend on the origin of the non-axisymmetric structures.
Rather, it is used to quantify the influence of a range
of generic spirals and hotspot geometries.

Using our formalism (with parameters
guided by a fit to the LFQPO properties of GRS1915+105) 
we expect that the RMS amplitude increases with disc inclination 
because the modulation from a spiral or blob
results from shadowing of the inner disc. 
Because the spiral is the hottest part of the  disc
for typical $r_c\sim 2r_{in}$,  we also find that the RMS amplitude should
increase with energy, except above energies where the
emission becomes strongly corona dominated.
More quantitative  modeling and more observations are needed.

If the time scale of a LFQPO in a black holes system
is indeed  determined by an orbit or pattern speed 
of a non-axisymmetric feature,
then at a given number of gravitational radii,
the LFQPO frequency should scale inversely  with the black hole mass.
If  accretion disks around AGN and microquasars incur  similar instabilities,
then a $1-10$Hz LFQPO in microquasars would correspond to a  $<10^{-6}$Hz
LFQPO in AGN.  Testing this prediction is presently difficult
because it pushes the present limits  of continuous 
observation times for individual AGN.

Finally note that although we have focused on black hole systems because of 
the LFQPO observations in microquasars, the analytic formalism for 
modeling the influence of spirals and hot spots herein 
does not depend on the compactness of the central object. It can 
also be applied to discs around 
neutron stars, white dwarfs, or young stellar objects.

{\refe More work is needed to incorporate our basic framework into detailed
models of individual LFQPO sourceds.}

{\bf Acknowledgments:}
We acknowledge support from
NSF grant AST-0406799 and NASA grant ATP04-0000-0016.
PV thanks  M. Tagger for discussions and we thank
M. Muno for discussions and allowing use of some code.

\newpage

\begin{table}[htbp]
\begin{tabular}{|c||c|c|c|c|c|c|c|c|c|}
\hline
\# &$h_o/r$ & $r_c/r_{in}$&$\alpha$&$\beta$&$\gamma$&$\delta$& angular
extend& rms \\
\hline
\hline
\# 1  & $0.01$  & $2.$  & $0.05$   & $0.5$    & $1$      &  $0.1$    &   $6 \pi$    &  2.3\% \\
\hline
\# 2  & $0.01$  & $2.$  & $0.05$   & $0.5$    & $1$      &  $0.1$    &   $4 \pi$   &  2.2\% \\
\hline
\# 3  & $0.01$  & $2.$  & $0.05$   & $0.5$    & $1$      &  $0.1$    &   $2 \pi$    &  2.2\% \\
\hline
\# 4  & $0.01$  & $2.$  & $0.05$   & $0.3$    & $1$      &  $0.1$    &   $2 \pi$    &  0.9\%  \\
\hline
\# 5  & $0.01$  & $2.$  & $0.05$   & $0.7$    & $1$      &  $0.1$    &   $2 \pi$    &  3.7\%  \\
\hline
\# 6  & $0.01$  & $2.$  & $0.05$   & $0.5$    & $2$      &  $0.1$    &   $2 \pi$    &  4.3\%\\
\hline
\# 7  & $0.01$  & $2.$  & $0.05$   & $0.5$    & $3$      &  $0.1$    &   $2 \pi$    &  7.3\%\\
\hline
\# 8  & $0.01$  & $2.$  & $0.05$   & $0.5$    & $2$      &  $0.2$    &   $2 \pi$    &  5.6\%\\
\hline
\# 9  & $0.01$  & $1.5$ & $0.05$   & $0.5$    & $2$      &  $0.1$    &   $2 \pi$    &  3.5\%\\
\hline
\# 10 & $0.01$  & $2.5$ & $0.05$   & $0.5$    & $2$      &  $0.1$    &   $2 \pi$    &  3.7\%\\
\hline
\# 11 & $0.01$  & $3.$  & $0.05$   & $0.5$    & $2$      &  $0.1$    &   $2 \pi$    &  3.2\%\\
\hline
\# 12 & $0.01$  & $2.$  & $0.03$   & $0.5$    & $2$      &  $0.1$    &   $2 \pi$    &  1.7\%\\
\hline
\# 13 & $0.01$  & $2.$  & $0.1$    & $0.5$    & $2$      &  $0.1$    &   $2 \pi$    & 12.3\%\\
\hline
\# 14 & $0.02$  & $2.$  & $0.05$   & $0.5$    & $2$      &  $0.1$    &   $2\pi$    &  8.6\%\\
\hline
\# 15 & $0.01$  & $2.$  & $0.05$   & $0.5$     & $2$      &  $0.1$    &   $2 \pi$    &  9.2\%\\
\hline
\# 16 & $0.01$  & $2.$  & $0.05$   & $0.5$    & $2$      &  $0.2$    &  $3\pi/2$ &  5.5\%\\
\hline
\# 17 & $0.01$  & $2.$  & $0.05$   & $0.5$    & $2$      &  $0.2$    &  $\pi$   &  4.7\%\\
\hline
\# 18 & $0.01$  & $2.$  & $0.05$   & $0.5$    & $2$      &  $0.2$    & $\pi/2$  &  3.0\%\\
\hline
\# 19 & $0.01$  & $2.$  & $0.05$   & $0.5$    & $2$      &  $0.2$    & $\pi/4$ & 1.6\%\\
\hline
\end{tabular}

\caption{For each simulation, all parameters, and the resulting RMS amplitude 
are given. The inclination angle for all cases
is $70^{\circ}$, which is the value for GRS1915+105 inferred from the 
orientation of the disk (Mirabel \& Rodr\'{\i}guez 1994), assuming the jet
is perpendicular to the jet. The angular extend of the
spiral/blob is given in fraction of $2\pi$.}
\label{tab:simul_modulation}
\end{table}

\begin{figure}[htbp]
\begin{tabular}{cc}
\epsfig{file=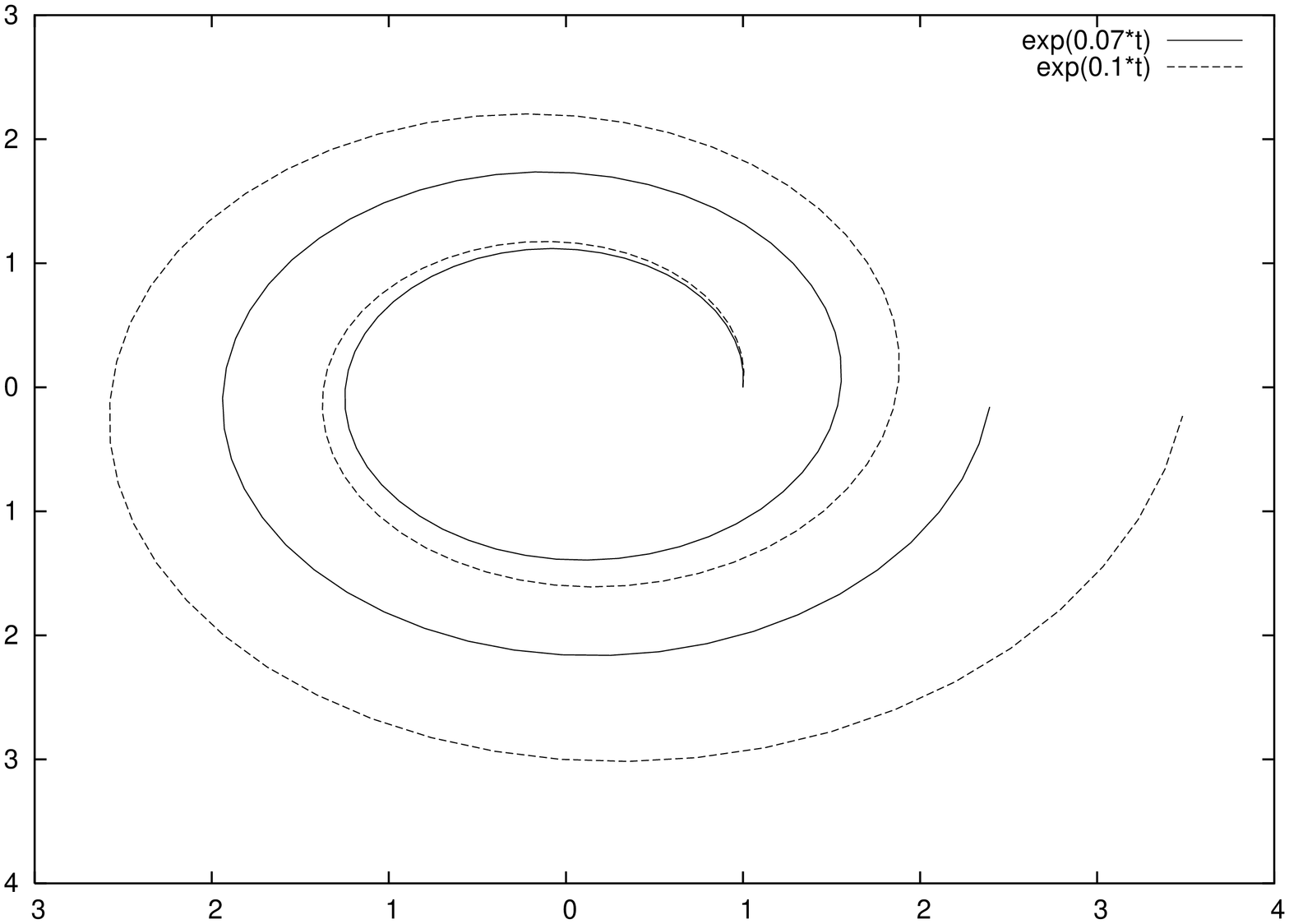,width =5cm,angle=-90}&  \\
\epsfig{file=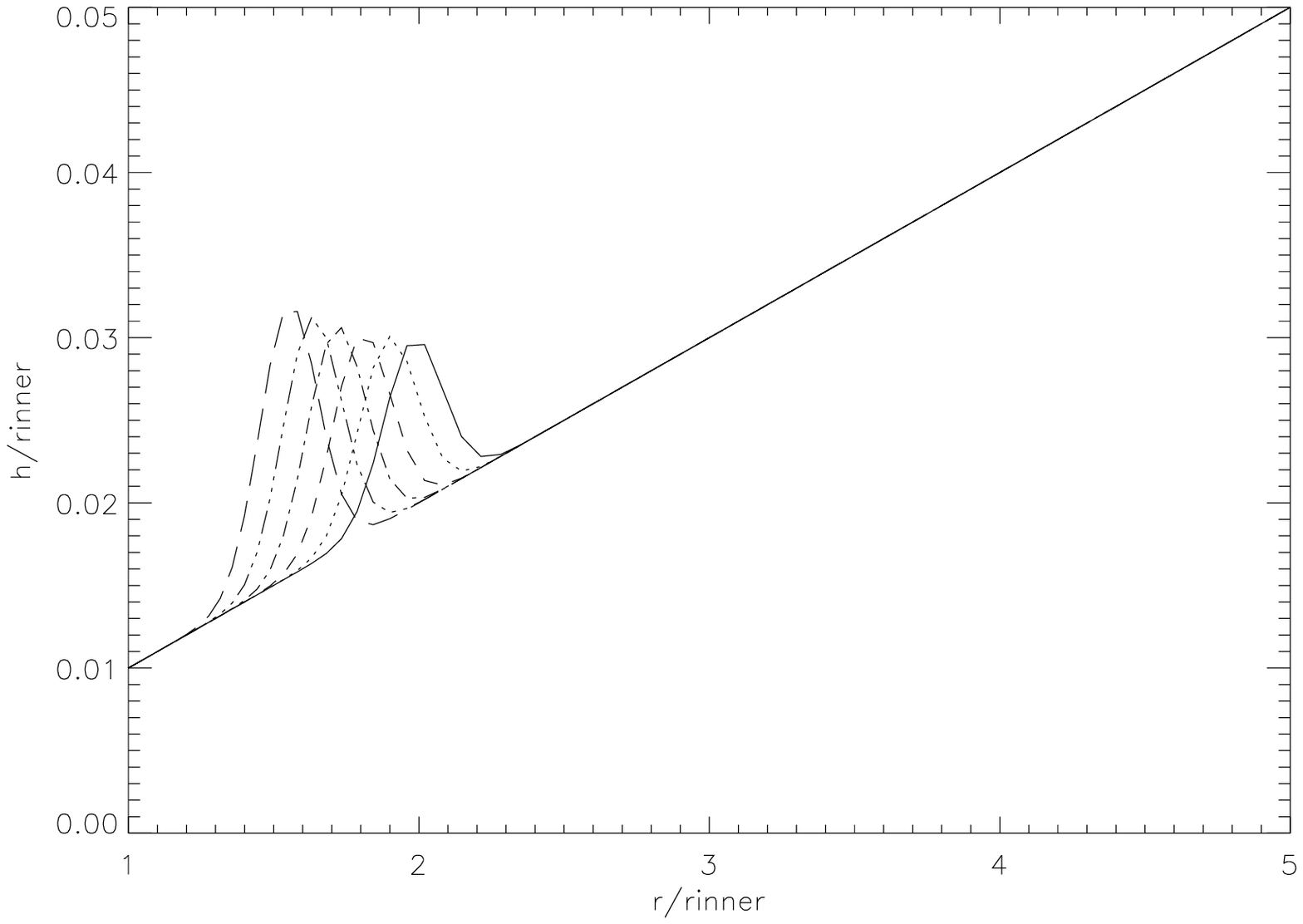,width=8cm}&
\epsfig{file=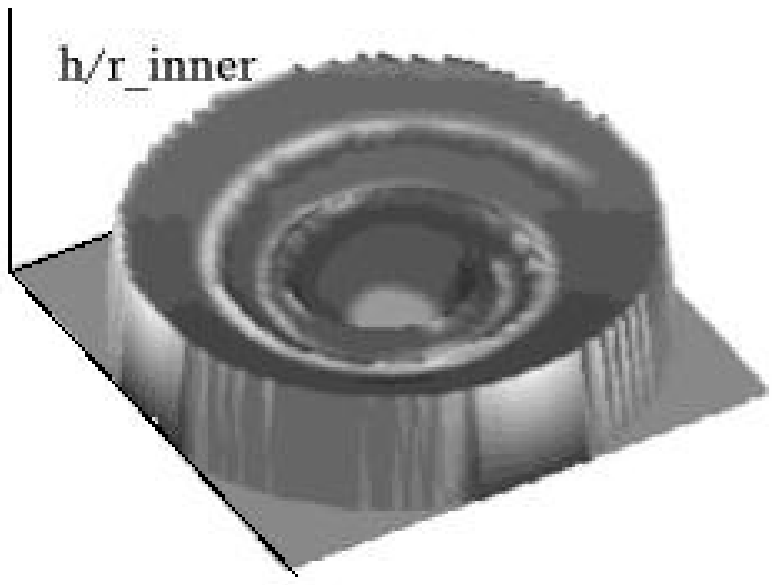}
\end{tabular}
\caption{
(a) Schematic top view of of spiral spines (line tracing the peak
in height of the spiral) for two different choices of $\alpha$.
(b) Schematic slices of the disc height profile for an example spiral 
as a function of radius for $h_o/r=0.01$. Profiles
at different azimuths are all projected into the same $r-z$ plane. 
The peak in the left-most curve 
occurs at $r_c$. The width of the curves is determined by
the parameter $\delta$. The downward slope of the line connecting the
peaks is determined by $\beta$ (for a fixed $\alpha$) and the
maxium height of the peaks is determined by $\tilde \gamma$ (see text).
(c) 3-D close up of the inner part of the disc with the spiral wave, 
the viewing angle is $70^\circ$ with 
$h_o/r = 0.01$, $r_c=3$, $\alpha = 0.07$, $\gamma = 0.3$,
$\beta = 2$, $\delta = 0.2$.}
\end{figure}

\begin{figure}
\epsffile{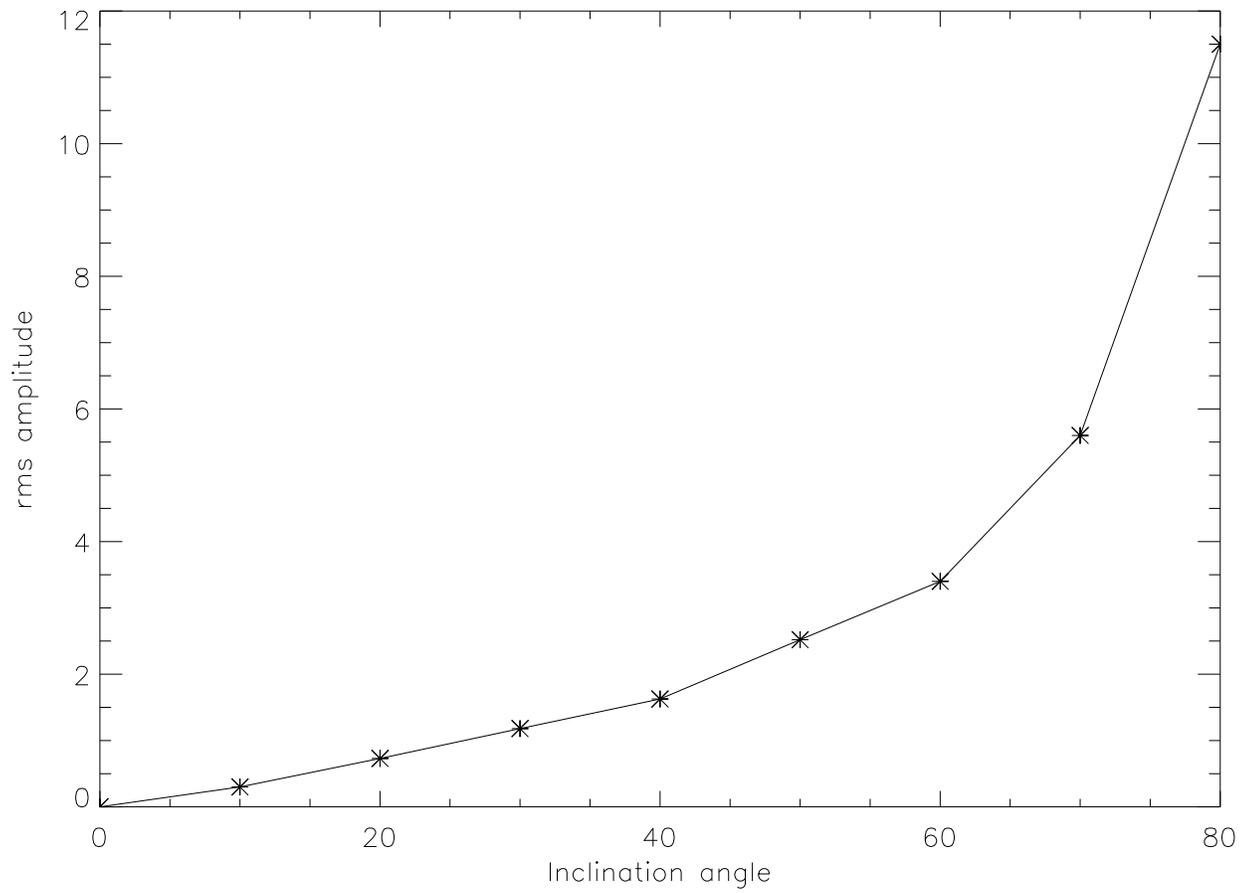}
\label{fig:incli}
\caption{ 
Evolution of the RMS as a function of the inclination angle $\theta$ 
for the simulation case \#8 of Table 1.}
\end{figure}

\end{document}